\documentclass[twocolumn,showpacs,pra]{revtex4}
\usepackage{amsmath,graphicx,bbm,natbib}

\begin{document}

\author{Felix Nissen}
\affiliation{Cavendish Laboratory, University of Cambridge,
  J.J.Thomson Ave., Cambridge CB3 0HE, UK}
\author{Jonathan Keeling}
\affiliation{Cavendish Laboratory, University of Cambridge,
  J.J.Thomson Ave., Cambridge CB3 0HE, UK}

\title{WKB approach and quantum corrections to  classical dynamics
  in the Josephson problem}
\begin{abstract}
  We apply a many-body WKB approach to determine the leading quantum
  corrections to the semiclassical dynamics of the Josephson model,
  describing interacting bosons able to tunnel between two localised
  states. The semiclassical dynamics is known to divide between
  regular oscillations, and self-trapped oscillations where the sign
  of the imbalance remains fixed. In both cases, the WKB wavefunctions
  are to be matched to Airy functions, yielding a modified
  Bohr-Sommerfeld quantisation condition. At the critical energy
  dividing normal and self-trapped oscillations, the WKB
    wavefunctions should instead be matched to parabolic cylinder
  functions, leading to a quantisation formula that is not just
    the Bohr-Sommerfeld formula, and recovering the known logarithmic
  quantum break times at this energy. This work thus provides another
  illustration of the usefulness of the WKB approach in certain
  many-body problems.

\end{abstract}
\maketitle

\section{Introduction}

The collective dynamics of a large number of interacting quantum
systems can often be described semiclassically, as mean-field
approximations of the dynamics of such systems become more accurate with increasing system size.  The great progress in trapping and
manipulating cold atoms, and in producing strong coupling between
confined photon modes and matter degrees of freedom have both led to
an increasing variety of systems where it becomes possible to study
many-body dynamics in isolated systems, and to investigate the extent
to which semiclassical descriptions are applicable.  In many cases,
the classical dynamics can show or has shown interesting collective
oscillations; examples include interconversion between fermionic atoms
and molecules\cite{andreev04,barankov04,yuzbashyan05:prb}, dynamical
superradiance in coupled light-matter
systems\cite{bonifacio70,eastham09}, optomechanical
oscillations\cite{brennecke08}, and the topic of this article,
Josephson oscillations between atoms trapped in different
wells\cite{milburn97,smerzi97,albiez05,levy07,esteve08}.

Josephson oscillations\cite{josephson62} of atoms between two
trapped condensates are well described by the Hamiltonian
\begin{equation}
  \label{eq:1}
  H=U(a^\dagger a - b^\dagger b)^2 + J (a^\dagger b + b^\dagger a).
\end{equation}
The semiclassical dynamics of this system have been studied in the context of the Lipkin-Meshkov-Glick Hamiltonian: Introducing the quantum operators $S_z=(b^\dagger b - a^\dagger a)/2, S_x=(a^\dagger b + b^\dagger a)/2, S_y=i(a^\dagger b - b^\dagger a) /2$, the Hamiltonian becomes 
 \begin{equation}
  \label{eq:LMG1}
  H=4 U S_z^2 + 2 J S_x .
\end{equation}
The semiclassical dynamics divide into two regimes: Small regular
oscillations between the wells for low energies and self-trapped oscillations, where the imbalance remains of a fixed sign
for high energies~\cite{leggett01}. Specifically, if we replace
the operators by c-numbers, the canonical equations of motion reduce
to the single expression
\begin{equation}
  \label{eq:2}
  \dot{S_z}^2 = 
  J^2N^2 - E^2 + 
  (2 U E - J^2)(2S_z)^2  -
  U^2 (2S_z)^4.
\end{equation}
Here we have used conservation of energy and particle number $N^2/4 =
| S |^2$. If $J^2N^2 -E^2 < 0$ and $2U E-J^2>0$ then the region of
allowed $S_z$ divides into two, excluding small population imbalances
(small values of $S_z$).  This means that for $2 N U > J$, then there
are two classes of oscillations; those with energies $E>NJ$ are
self-trapped oscillations\cite{smerzi97,milburn97,leggett01} where the
sign of imbalance remains fixed, and for those with $E<NJ$ the sign of
the imbalance varies periodically. Both regimes have been observed
experimentally \cite{albiez05,levy07}. The Lipkin-Meshkov-Glick
Hamiltonian, Eq.~(\ref{eq:LMG1}), has been studied in a coherent
state representation semiclassically \cite{vidal07} and with quantum
corrections \cite{vidal08}. It has also been studied for small numbers
of particles where quantum corrections may play a larger role
\cite{tonel05}.

Instead of the spin representation, this article maps the Josephson
problem onto a discrete Schr\"odinger equation. As
  this is a one-dimensional problem, it can be approximately solved by
  the WKB approach~\cite{landauQM}.  Expanding for large system size
$N$, the WKB wavefunctions can be found for the allowed and forbidden
regions of Fock space. At the boundaries, these are matched to exact
solutions of the Schr\"odinger equation, which often yields Airy
functions. In such cases the spectrum is given by the usual
Bohr-Sommerfeld quantisation condition on the action $A(E_n) =
\int p\, \mathrm{d}x = 2\pi (n+1/2)$ at a particular energy level
$E_n$. Using this phase-space quantisation approach, the Josephson
problem has been studied in \cite{graefe07,trippenbach}.  One notable feature
  of the validity of semiclassics for the Josephson problem is that
  the quantum break times --- the characteristic time at which quantum
  and classical dynamics start to differ ---may grow only
  logarithmically, rather than algebraically, with $N$ when near the
  energy dividing self-trapped and non self-trapped oscillations.
  This has been seen both by using a cumulant expansion
\cite{vardi01,anglin01} and by phase-space quantisation
\cite{boukobza}.  The existence of an associated logarithmic
divergence in the density of states has been discussed by
  \citet{hooley04} for the related model of a single particle moving
in an infinite tight-binding lattice with an overall harmonic trap.

The aim of this article is to apply the many-body WKB approach as
discussed in Ref.\cite{babelon09} to the Josephson problem of
Eq.~(\ref{eq:1}), and to use this to find the quantum break times for
critical and non-critical levels.  We solve the Schr\"odinger equation
for large system size $N$ to next-to-leading order.  The WKB
  wavefunctions in the allowed and forbidden regions are then matched
  to appropriate special functions at the boundaries.  For most
  energies, the appropriate special function is the Airy function,
  which then recovers the Bohr-Sommerfeld quantisation condition
  derived in \cite{graefe07,trippenbach}.  However, near the critical
  energy separating self-trapped and non self-trapped oscillations,
  the relevant special functions are instead parabolic cylinder
  functions, and these lead to a different quantisation condition,
  that goes beyond the Bohr-Sommerfeld formula of
  \cite{graefe07,trippenbach}.  Even away from the critical level,
the full WKB analysis yields information beyond Bohr-Sommerfeld
quantisation: The allowed region divides into a regime where matching
to a purely decaying Airy function is appropriate and one where the
decaying solution is also highly oscillatory. While this behaviour
modifies the nature of the wavefunctions, its effect on the
  quantisation condition is only a jump in the Maslov index. As such, it
leaves the density of states unaffected, and so it was not crucial to
previous work in the context of Bohr-Sommerfeld quantisation.

The approach presented in this article has also been used in
Ref. \cite{babelon09} to study the Tavis-Cummings model \cite{tavis68}
describing e.g. the dynamics of two-level systems coupled to
photons. The many-body WKB approach gives quantum dynamics for this
model very similar to the classical dynamics for most energy levels
and initial conditions. However, there are critical levels, for which
the matching of the WKB wavefunctions becomes more complex and gives
rise to anharmonicity scaling as the logarithm of system size
\cite{keeling09:tavis-cummings}.  In both the Tavis-Cummings
  model, and in the Josephson model, these critical levels show
logarithmic quantum break times, analogous to the $\ln(\hbar)$
quantum break times found near unstable classical states in
single particle quantum problems~\cite{casati_chirikov}.  A
further class of problems in which this many-body WKB approach may be
of use concerns variations of problems such as the Josephson problem,
and the Tavis-Cummings model, in which parameters are varied as a
function of time to give many-body generalisations of Landau-Zener
problems\cite{liu02,zobay00,pazy05,witthaut06,altland08,itin09}.  Many
treatments of this problem have been effectively semiclassical, and
the WKB approach may provide a method to determine whether large
quantum corrections can ever arise as a result of transitions to and
from critical levels.  An approach along these lines has been
  explored by \citet{witthaut06}.

The remainder of this article is arranged as follows.
Section~\ref{sec:discr-wkb-appr} maps the many-body Hamiltonian to a
form amenable to solution by the WKB method, and then provides the
ingredients necessary to determine this solution. These ingredients
are the WKB wavefunctions in the classically allowed and forbidden
regions, given in Sec.~\ref{sec:wkb-wavef-allow}, and the connection
formula holding at the boundaries between these regions, given in
Sec.~\ref{sec:connection-formulae}.
Section~\ref{sec:bound-match-quant} then combines these ingredients to
give the resulting quantisation conditions, which are specified in
three separate ranges of energies, according to whether one is above,
near, or below the critical energy level (the energy level dividing
self-trapped and non-self-trapped oscillations).  From these
quantisation conditions, section~\ref{sec:conclusions} then extracts
the scaling of the quantum break time as a function of system
size, and summarises the results found.

\section{Discrete WKB approach to Josephson equation}
\label{sec:discr-wkb-appr}

To use the WKB approach, we need to produce a discrete Schr\"odinger
equation, by writing the state of the system in terms of the total
number of particles $N$, and the imbalance $n$:
\begin{equation}
  \label{eq:3}
  | \Psi \rangle = \sum_{n=-N}^N \psi_n
  \frac{a^{\dagger(N+n)/2} b^{\dagger(N-n)/2}}{%
    \sqrt{[(N+n)/2]![(N-n)/2]!}}
    |\Omega \rangle
\end{equation}
where $n$ is restricted to the same odd/even parity as $N$.  Acting on
this state with the many-body Hamiltonian, and looking for eigenstates
with energy $E$, yields a discrete Schr\"odinger equation:
\begin{multline}
  \label{eq:4}
  (E -  U n^2) \psi_n = \frac{J}{2} \left[
    \sqrt{(N+n)(N-n+2)} \psi_{n-2}
    \right.\\\left.+
    \sqrt{(N+n+2)(N-n)} \psi_{n+2}
    \right].
\end{multline}

Writing $E=\epsilon J N, U=J u/N, z=n/N$, one can separate the system
size dependence from the other parameter dependence and write:
\begin{multline}
\label{eq:5}
  (\epsilon - u z^2) \psi(z) =
  \frac{1}{2} 
  \sqrt{\left(1+z\right)\left(1 - z +\frac{2}{N} \right)}
  \psi\left(z - \frac{2}{N} \right)
  \\+
  \frac{1}{2}
  \sqrt{\left(1-z\right)\left(1 + z +\frac{2}{N} \right)}
  \psi\left(z + \frac{2}{N} \right).
\end{multline}
In these units, self-trapped states exist only if $u>1/2$ and
$\epsilon>1$.

\subsection{WKB wavefunction}
\label{sec:wkb-wavef-allow}
For a given energy, one may divide the range of $-1<z<1$ into
classically allowed and forbidden regions.  These are distinguished by
oscillating versus decaying wavefunctions, and correspond directly to
the regions of $S_z=Nz$ in Eq.~(\ref{eq:2}) which the classical dynamics
explores.
\subsubsection{Allowed region}
\label{sec:allowed-region}
\begin{figure}[htpb]
  \centering
  \includegraphics[width=3.2in]{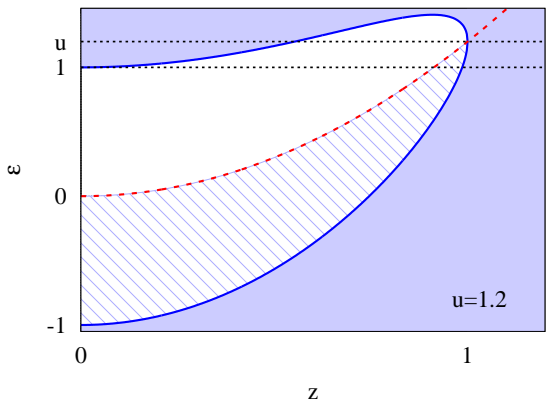}
  \caption{(Color online) Boundaries between allowed and
    forbidden regions, and between regions which needs extra factor of
    $(-1)^{n/2}$.  Solid shaded region is classically forbidden,
    hatched region ($\epsilon < u z^2$) requires extra factor of
    $(-1)^{n/2}$ to allow matching to connection formula at
    boundary. Plotted for $u=1.2$}
  \label{fig:allowed-forbidden}
\end{figure}

% Write down WKB forms, and associated phase in allowed region
% Discuss boundary of allowed region, and refer to figure
% Discuss
  In the allowed region, the WKB ansatz has the form:
\begin{equation}
  \label{eq:6}
  \psi(z) = b(z) \left[
    C_+ e^{i (N W_0 + W_1)} +
    C_- e^{-i (N W_0 + W_1)} 
  \right].
\end{equation}
where $W_0$ and $W_1$ are the $z$-dependent phase terms at leading and
next-to-leading order.  By substituting this into Eq.~(\ref{eq:5}) and
identifying real and imaginary terms of the same order in $1/N$, one
finds the definitions:
\begin{align}
  \label{eq:7}
  \cos(2 W_0^\prime(z))
  &=
  \frac{\epsilon - u z^2}{\sqrt{1 - z^2}}
  \\
  \label{eq:8}
  b(z) 
  &=
  \left[ 1 - z^2 - (\epsilon - u z^2)^2 \right]^{-1/4}
  \\
  \label{eq:9}
  W_1^\prime(z) &= \frac{\epsilon - u z^2}{%
    2 (1-z^2) \sqrt{1 - z^2 - (\epsilon - u z^2)^2}}.
\end{align}
(Note that the sign written for $W_1^\prime$ assumes that the solution of
Eq.~(\ref{eq:7}) is taken such that $\sin(2 W_0^\prime) \geq 0$).  For
this solution to be valid, it is clearly necessary that $|\epsilon-u
z^2| \leq \sqrt{1-z^2}$, which defines the classically allowed region
shown in Fig.~\ref{fig:allowed-forbidden}.  

Within the classically allowed region, there is also a division
between $\epsilon > u z^2$ for which $0\leq 2 W_0^\prime < \pi/2$, and
$\epsilon < u z^2$ for which $\pi/2 < 2 W_0^\prime \leq \pi$.  While
this distinction is unimportant within the allowed region, it will
later be necessary to distinguish these regions, so that the
connection formulae at the boundary of the allowed region can be
written in terms of smooth wavefunctions.  If $\epsilon > u z^2$, then
at the boundary $W_0^\prime \to 0$, so the wavefunction is smooth.
For $\epsilon < u z^2$, $W_0^\prime \to \pi/2$, which indicates that
$N W_0 = i N x\pi/2$, which is rapidly varying.  This rapid variation
can be removed in such cases by instead defining $\psi_n \to \psi_n
(-1)^{n/2}$. Since only values of $n$ with the same parity as $N$
exist, this corresponds to terms having alternating signs.  After this
transformation, Eq.~(\ref{eq:7}--\ref{eq:9}) are modified by replacing
$(\epsilon - u z^2) \to -(\epsilon - u z^2)$.  The regions where this
transformation is necessary are indicated in
Fig.~\ref{fig:allowed-forbidden} by hatching.  We will use the
notation $\tilde{W}_0, \tilde{W}_1$ for the phases calculated with
this additional minus sign.  When part of the allowed region does
require this transformation, and part does not (i.e.\ for
$0<\epsilon<u$), it is necessary to consider carefully the connection
between these regions; this is discussed further in
Sec.~\ref{sec:conn-with-allow}.

\subsubsection{Forbidden region}
\label{sec:forbidden-region}
\begin{figure}[htpb]
  \centering
  \includegraphics[width=3in]{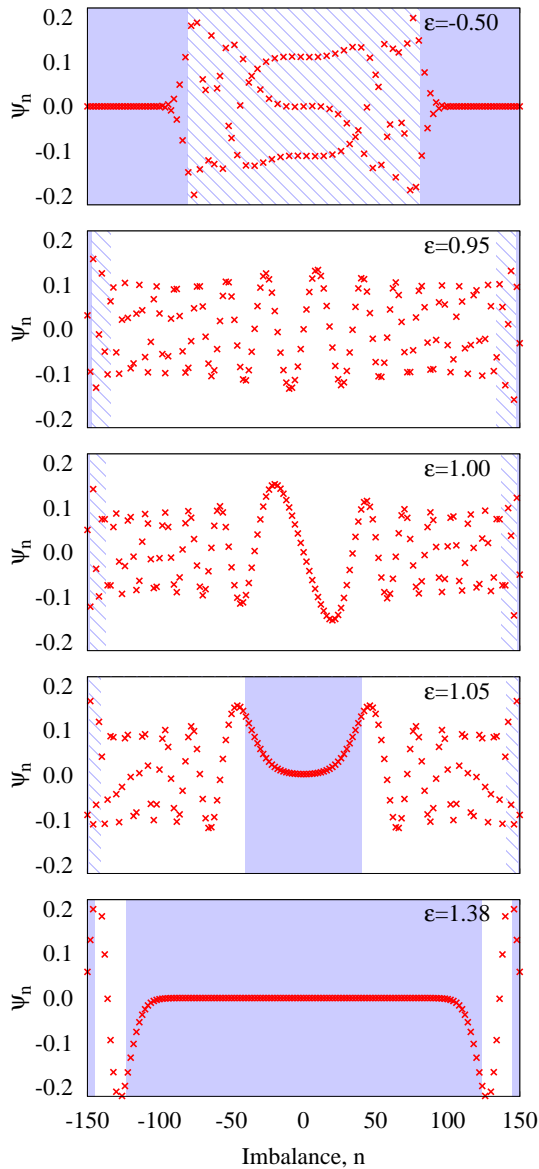}
  \caption{(Color online) Wavefunctions at various
    characteristic energies, showing behaviour in allowed and
    forbidden regions.  Background shading is as in
    Fig.\ref{fig:allowed-forbidden}. Plotted for $u=1.2$}
  \label{fig:wavefunctions}
\end{figure}

% Talk about form in forbidden region
In the forbidden region, the wavefunction exponentially decays, so
the WKB ansatz becomes:
\begin{equation}
  \label{eq:10}
  \psi(z) = b(z) \left[
    C_+ e^{(N \Omega_0 + \Omega_1)} +
    C_- e^{-(N \Omega_0 + \Omega_1)} 
  \right].
\end{equation}
In this case, there is no strict distinction between the terms to be
incorporated in $b(z)$ and $\Omega_1(z)$ --- both describe the real
part at order $1/N$.  However, it is convenient for the connection
formulae to keep $b(z)$ as in Eq.~(\ref{eq:8}), after which
$\Omega_{0,1}$ can be identified by powers of $1/N$ as:
\begin{align}
\label{eq:11}
  \cosh(2 \Omega_0^\prime)
  &=
  \frac{\epsilon - u z^2}{\sqrt{1 - z^2}}
  \\
\label{eq:12}
  \Omega_1^\prime &=\frac{-(\epsilon - u z^2)}{%
    2 (1-z^2) \sqrt{(\epsilon - u z^2)^2} - 1 + z^2 }
\end{align}
In this case, the sign of $\Omega_1^\prime$ assumes $\Omega_0^\prime
\geq 0$.  In the forbidden region at small $z$ one always has
$\epsilon > u z^2$ so no alternating sign factors are needed, but in
the forbidden region at large $z$, a factor $(-1)^{n/2}$ may be needed
so that $\cosh(2 \Omega_0^\prime) > 0$.  The distinction between
exponential decay, and exponential decay with alternating signs is
clearly visible in the exact wavefunctions shown in
Fig.~\ref{fig:wavefunctions}.  In the forbidden region at small $z$,
one will have $|C_-|=|C_+|$ (assuming that $\Omega$ is measured from
$z=0$), since the wavefunctions will be either odd or even functions
of $z$.  In the region at large $z$, only one of $C_\pm$ will be
non-zero, describing exponential decay for $z\to \mp\infty$.

\subsubsection{Critical level}
\label{sec:critical-behaviour}
% Discus existence of critical level
At the critical level $\epsilon=1$, there is a
bifurcation, and the WKB wavefunction in the allowed region must be
matched to a different special function for $z \simeq 0$. For this purpose it is convenient to
write $\epsilon=1 + \lambda/N$, where $\lambda$ is small compared to
$N$, and to rewrite the WKB form for the allowed regime making use of
this --- because the deviation of the energy from $1$ is now considered
as being of order $1/N$, its effect is relegated from $W_0$ to $W_1$,
\begin{align}
  \label{eq:19}
  \cos(2 W_0^\prime(z))
  &=
  \frac{1 - u z^2}{\sqrt{1 - z^2}}
  \\
  \label{eq:20}
  b(z) 
  &=
  \left[z^2\left(2u  - 1    - u^2 z^2\right) \right]^{-1/4}
  \\
  \label{eq:21}
  W_1^\prime(z) &= \frac{1- \lambda + (\lambda- u) z^2}{%
    2 (1-z^2) |z| \sqrt{2u  - 1    - u^2 z^2}}.
\end{align}
One can see from Eq.~(\ref{eq:21}) that $W_1$ will be logarithmically
divergent as $z \to 0$.  The form of $\psi(z)$ for $z\simeq 0$ will
give the natural cutoff to this logarithm, which will be found to
depend on the system size $N$, so exactly at this critical level, the
eigenvalue spacing will depend logarithmically on the system size,
which in turn leads to logarithmic dependence of the quantum break
time on system size.

\subsection{Connection formulae}
\label{sec:connection-formulae}

% Write equation in general, note need for smoothness of function
To connect the WKB ansatz valid in the classically allowed and
classically forbidden regions, one needs the solution valid near the
boundary.  If the wavefunction is smooth near this point, one may
expand Eq.~(\ref{eq:4}) directly in powers of $1/N$ to give a
differential equation for the wavefunction.  (Note that in the WKB
ansatz, there was no such assumption of smoothness, since we did not
require $W_0^\prime$ to be small.)  To ensure the wavefunction is
smooth, a factor $(-1)^{n/2}$ may be needed according to whether the
boundary is in the hatched on unhatched region of
Fig.~\ref{fig:allowed-forbidden}, as discussed above.  Both these
cases can be considered together, by replacing $(\epsilon - uz^2) \to
|\epsilon - uz^2|$ to take account of the two possible signs.  With
this change, expanding Eq.~(\ref{eq:5}) to order $1/N^2$ one finds:
\begin{multline} \label{eq:13}
  |\epsilon - u z^2| \psi
  =
  \left[\sqrt{1-z^2} + \frac{1}{N\sqrt{1-z^2}}\right] \psi
  \\+
  \frac{2\sqrt{1-z^2}}{N^2} 
  \left[ \psi^{\prime\prime} - \frac{z \psi^\prime}{1-z^2} 
    + (\ldots) \psi
  \right] + \mathcal{O}\left(N^{-3}\right)
\end{multline}
When considering the leading and next to leading order behaviour
(i.e.\ classical behaviour plus leading order quantum corrections), the
term of order $N^{-2}$ term involving $\psi$ may be neglected, as
there are larger terms involving $\psi$, whereas the terms involving
$\psi^\prime$ and $\psi^{\prime\prime}$ should be kept.  Rewriting the
above equation one has:
\begin{equation}
  \label{eq:14}
  0 = \psi^{\prime\prime}
  - \frac{z}{1-z^2} \psi^\prime
  + \left[
    \frac{N}{2(1-z^2)}
    + \frac{N^2}{2}\!
    \left(1 - \frac{|\epsilon - u z^2|}{\sqrt{1-z^2}} \right)
  \right]\psi.
\end{equation}

\subsubsection{Regular boundaries --- Airy functions}
\label{sec:regul-bound-airy}

Away from the critical level $\epsilon=1$, the boundary between the
allowed and forbidden region is given by $|\epsilon-uz^2| = \sqrt{1-z^2}$, 
which has the outer (inner) solutions:
\begin{equation}
  \label{eq:15}
  z_{o,i}^2 = \frac{1}{2 u^2} \left[
    2 u \epsilon - 1 \pm \sqrt{1 - 4 u \epsilon + 4 u^2} 
  \right]
\end{equation}
% Write equation for a normal boundary, and note solution
Near these boundaries, we want to write Eq.~(\ref{eq:14}) in terms of
distance $\zeta$ from the boundary, $z=\pm z_{o,i} + \zeta$.  Since
Eq.~(\ref{eq:14}) is even under $z \to - z$, it is clear that
the equation at $z=z_{o,i} + \zeta$ and $z=-z_{o,i} + \zeta$ are
related by $\zeta\to-\zeta$.  Therefore, considering the former of
these, we may write Eq.~(\ref{eq:14}) as:
\begin{equation}
  \label{eq:16}
  0 = \psi^{\prime\prime} - \alpha_{o,i}\psi^\prime + \left[
    \beta_{o,i} +  f_{o,i}(\zeta) \right] \psi
\end{equation}
with $\alpha_{o,i} = z_{o,i} /(1-z_{o,i}^2), \beta_{o,i} =
N/[2(1-z_{o,i}^2)]$ and:
\begin{align}
  \label{eq:17}
  f_{o,i}(\zeta) &\simeq
   \frac{N^2}{2} \left[
    1 - \frac{|\epsilon - u z_{o,i}^2-2u z_{o,i} \zeta|}{%
      \sqrt{1-z_{o,i}^2-2 z_{o,i} \zeta}} \right]
  \nonumber \\ &\simeq
  \frac{N^2}{2}z_{o,i} \zeta 
  \left[ \frac{2 u (\epsilon - u z_{o,i}^2) -1}{1-z_{o,i}^2} \right]
  \nonumber \\ &=
  \mp \frac{N^2}{2} 
  \frac{z_{o,i} \zeta \sqrt{1-4 u \epsilon + 4 u^2}}{1-z_{o,i}^2}
  \stackrel{\text{def}}{=} \gamma_{o,i} \zeta
\end{align}
In the above, we have assumed that $\zeta$ is small so the sign of
$\epsilon - u z^2$ does not change, and the last line has used the
form of $z_{o,i}^2$ in Eq.~(\ref{eq:15}), with the $\mp$ signs
corresponding to the outer (inner) boundary.  With $f(\zeta)=\gamma
\zeta$, the solution Eq.~(\ref{eq:16}) can be written using Airy
functions:
\begin{equation}
  \label{eq:18}
  \psi = e^{- {\alpha \zeta/}{2}}
  \left[ C_a Ai(-\gamma^{1/3} \xi) + C_b Bi(-\gamma^{1/3}\xi) \right]
\end{equation}
with $\xi = \zeta + \beta/\gamma - \alpha^2/4 \gamma$.  For the outer
boundaries we match to a decaying solution, so $C_b=0$.  Since
$\gamma_o$ is negative, the solutions is oscillatory for $\zeta<0$ and
decaying for $\zeta > 0$ as expected.  For the inner boundaries, both
exponentially decaying and growing parts are required. Since
$\gamma_i$ is positive, the solutions are oscillatory for $\zeta > 0$
and grow/decay for $\zeta <0$.

\subsubsection{Critical boundary --- Parabolic Cylinder functions}
\label{sec:crit-bound-parab}

% Write equation for a critical level
Near $\epsilon=1$, the inner boundary becomes an extremum at $z=0$,
and the form of Eq.~(\ref{eq:14}) is different to that in
Sec.~\ref{sec:regul-bound-airy}.  To study energies near this level,
we will write $\epsilon=1+\lambda/N$ as in
Sec.~\ref{sec:critical-behaviour}.  Since $\epsilon=1, z=0$ never
requires a factor $(-1)^{n/2}$ in the wavefunction, Eq.~(\ref{eq:14})
can always be written near this point as:
\begin{equation}
  \label{eq:22}
    0 = \psi^{\prime\prime}
    - z \psi^\prime
    + \left[
      \frac{N}{2}(1-\lambda)
      + \frac{N^2}{4} (2u-1) z^2
  \right]\psi.
\end{equation}
To solve this equation, it is convenient to define:
\begin{equation}
  \label{eq:24}
  \mu = \sqrt{2u - 1}, \qquad \chi=\frac{1-\lambda}{2\mu}.
\end{equation}
After removing a Gaussian factor $\psi = e^{z^2/4} f$, this can be
recognised as Weber's equation\cite{whittaker_watson}; in terms of
$\xi = z e^{-i\pi/4} \sqrt{N \mu}$ one has $0 = f^{\prime\prime} +
(i \chi - \xi^2/4) f$ with solutions in terms of parabolic cylinder
functions.  The general solution can be written as:
\begin{multline}
    \psi =
    e^{z^2/4} \left[
      \alpha D_{i\chi - 1/2}\left( e^{-i \pi/4} \sqrt{N \mu} z \right)
    \right.
    \\\left.+
      \beta D_{i\chi - 1/2}\left( e^{3 i \pi/4} \sqrt{N \mu} z \right)
    \right].
\end{multline}
The Gaussian prefactor only contributes at order $1/N$, so in matching
the asymptotics of $N W_0, N\Omega_0$ to $\psi$, this prefactor can
be dropped.  It is clear that for $z \to - z$, this expression changes
as $\alpha \leftrightarrow \beta$.  The asymptotic expansion of this
expression for large $x$ given in Ref.~\cite{whittaker_watson} can be
written:
\begin{align}
  \label{eq:25}
  \psi &=
  \left[
    \alpha \exp\left(
      \frac{\pi \chi}{4} + i \frac{\pi}{8}
    \right)
    +
    \beta \exp\left(
      - \frac{3 \pi \chi}{4} - i \frac{3 \pi}{8}
    \right)
  \right]
  \nonumber\\
  &\quad\times \exp\left(
    i\frac{N \mu z^2}{4}
    +
    \left[i \chi - \frac{1}{2} \right]
    \ln \left[ z \sqrt{N \mu} \right]
  \right)
  \nonumber\\
  &+ \beta
  \frac{\sqrt{2\pi}}{\Gamma[1/2 - i \chi]}
  \exp\left(
    -\frac{\pi \chi}{4} + i \frac{\pi}{8}
  \right)
  \nonumber\\
  &\quad\times \exp\left(
    -i\frac{N \mu z^2}{4}
    +
    \left[-i \chi - \frac{1}{2} \right]
    \ln \left[ z \sqrt{N \mu} \right]
  \right).
\end{align}
As noted in Sec.~\ref{sec:critical-behaviour}, the phase has a
logarithmic divergence at small $z$, and the form of the wavefunction
given here provides the cutoff for the logarithm, $1/\sqrt{N \mu}$,
which depends on system size.

\subsubsection{Connection within allowed region}
\label{sec:conn-with-allow}

% Note that particular problem is when we match to two different types
% of boundary
In addition to connection formulae at the boundaries of allowed and
forbidden, for energies in the range $1<\epsilon<u$, it is necessary
to connect solutions with and without the extra factor of $(-1)^{n/2}$
in the middle of the allowed region.  

In such cases, we may write:
\begin{equation}
  \label{eq:23}
  \psi = 
  \begin{cases}
    \phantom{(-1)^{n/2}}
    \left[
      C_+ e^{i W} + C_- e^{-iW}
    \right] & |z|<z_s
    \\[0.1em]
    (-1)^{n/2}
    \left[
      \tilde{C}_+ e^{i \tilde{W}} + 
      \tilde{C}_- e^{-i \tilde{W}}
    \right] & |z| > z_s
  \end{cases},
\end{equation}
where $z_s=\sqrt{\epsilon/u}$ is the point at which the sign of
$\epsilon- uz^2$ changes.  The question is how to relate $C_{\pm}$ to
$\tilde{C}_{\pm}$ at the boundary.  To fully define the the phases $W,
\tilde{W}$ one must specify the limits of integration.  It is
convenient to choose:
\begin{equation}
  \label{eq:26}
  W(z) = \int_{z_i}^z \!\!dz [N W^\prime_0 + W^\prime_1],
  \quad
  \tilde{W}(z) = \int_{z_o}^z \!\!dz [N \tilde{W}^\prime_0 + \tilde{W}^\prime_1]
\end{equation}
and to then define the phases at $z_s$ as $\Delta W_{\text{in}} =
W(z_s), \Delta \tilde{W}_{\text{out}} = - \tilde{W}(z_s)$.  To
determine the connection, we focus on the phase due to $W_0^\prime$;
expanding for $z=z_s + \zeta$ one has:
\begin{equation}
  \label{eq:27}
  \cos(2 W_0^\prime) = \frac{-2 u\sqrt{\epsilon } \zeta}{\sqrt{u-\epsilon}}
  = -\cos(2 \tilde{W}_0^\prime),
\end{equation}
and integrating one has:
\begin{align}
  \label{eq:78}
  W(z_s+\zeta) &= +\Delta W_{\text{in}}
  + N \left[
    \frac{\pi}{4} \zeta + 
    \frac{\sqrt{\epsilon u} }{\sqrt{1-\epsilon/u}} \frac{\zeta^2}{2}
  \right]
  \\
  \label{eq:80}
  \tilde{W}(z_s+\zeta) &= -\Delta \tilde{W}_{\text{out}}
  + N \left[
    \frac{\pi}{4} \zeta -
    \frac{\sqrt{\epsilon u} }{\sqrt{1-\epsilon/u}} \frac{\zeta^2}{2}
  \right].
\end{align}
To match the $\zeta$ dependence near $\zeta=0$ one can write the
prefactor $(-1)^{n/2} = \exp[i \pi N (z_s + \zeta)/2]$ which gives the
matching conditions:
\begin{equation}
  \label{eq:28}
  C_- = 
  \tilde{C}_+
  \exp\left[
    i \Delta {W}_{\text{in}}
    - i \Delta \tilde{W}_{\text{out}}
    - i \frac{\pi N}{2} z_s
  \right],
\end{equation}
and the complex conjugate for the other pair.

\section{Boundary matching and quantisation conditions}
\label{sec:bound-match-quant}

\begin{figure}[htpb]
  \centering
  \includegraphics[width=3.2in]{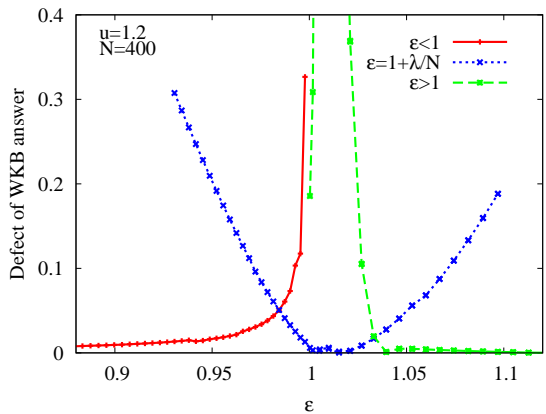}
  \caption{(Color online) Accuracy of WKB-derived quantisation
    condition.  Each line corresponds to difference of appropriately
    defined phase difference, and nearest integer multiple of $\pi$,
    using quantisation conditions appropriate to energies below, near
    and at the critical level. Plotted for $u=1.2$}
  \label{fig:wkb-accuracy}
\end{figure}

By connecting the wavefunctions in the allowed and forbidden regions to
the Airy or parabolic cylinder functions at the boundaries, one can
derive conditions relating the coefficients $C_{\pm}$ in these various
regions.  In order to simultaneously satisfy all the equations, it is
necessary to have certain conditions on the WKB phases; this provides
quantisation conditions on the allowed energies.  This section will
derive these quantisation conditions in the three cases $\epsilon<1,
\epsilon>1$ and $\epsilon=1+\lambda/N$.  In all cases the quantisation
condition can be reduced to a form $f(\epsilon) = m \pi$, where $m$ is
an integer.  Figure~\ref{fig:wkb-accuracy} illustrates the accuracy of
these WKB quantisation methods by plotting $\min_m(f(\epsilon_n) - m
\pi)$, where $\epsilon_n$ is an exact eigenvalue --- i.e.\ determining
the extent to which the exact eigenvalues obey the WKB quantisation
condition.

\subsection{Below critical level}
\label{sec:below-crit-energy}

We consider separately the cases $\epsilon<0$ and $0<\epsilon<1$.  For
$\epsilon<0$, the quantisation condition is simple as the factor
$(-1)^{n/2}$ is needed throughout the allowed region.  Let us define
the limits of integration for $W(z)$ such that:
\begin{equation}
  \label{eq:29}
  \tilde{W}(z) 
  = \int_{0}^z \!\!dz [N \tilde{W}^\prime_0 + \tilde{W}^\prime_1], \qquad
  \Delta \tilde{W} = \tilde{W}(z_o).
\end{equation}
Then, by expanding the equation for $\tilde{W}_0^\prime$
(i.e.\ Eq.~(\ref{eq:19}) with an extra minus sign) near $z=z_0+\zeta$, and
comparing the result to the last line of Eq.~(\ref{eq:17}) one can show that:
\begin{equation}
  \label{eq:30}
  1 - 2 (\tilde{W}_0^\prime)^2 
  = 
  1- \left[\frac{2 u (\epsilon-u z_o^2)-1}{1-z_o^2} \right] z_o \zeta
  =
  1 - \frac{2 \gamma_o}{N^2} \zeta.
\end{equation}
Remembering that $\gamma_o<0$, integrating the expression for
$\tilde{W}_0^\prime$ gives:
\begin{equation}
  \label{eq:31}
  \tilde{W}(z_o + \zeta) = \Delta\tilde{W} - \frac{2}{3} |\gamma_o|^{1/2}
  |\zeta|^{3/2}, \qquad (\zeta<0).
\end{equation}
In order to match $\psi = (-1)^{n/2} [\tilde{C}_+\exp(i \tilde{W}) +
\tilde{C}_-\exp(-i\tilde{W})]$ to the expansion of the Airy
function\cite{whittaker_watson} appearing in Eq.~(\ref{eq:18}):
\begin{equation}
  \label{eq:32}
  \psi 
  \propto
  Ai[|\gamma_o|^{1/3} (\zeta + \ldots)] 
  \simeq 
  \sin\left(\frac{\pi}{4} + \frac{2}{3} |\gamma_o|^{1/2} |\zeta|^{3/2} \right)
\end{equation}
gives the condition $(\tilde{C}_+/\tilde{C}_-) \exp(2i \Delta
\tilde{W}) = i$.  Repeating this procedure at $z=-z_o+\zeta$, one has
instead
\begin{equation}
  \label{eq:31}
  \tilde{W}(-z_o + \zeta) = -\Delta\tilde{W} + \frac{2}{3} |\gamma_o|^{1/2}
  |\zeta|^{3/2}, \qquad (\zeta>0),
\end{equation}
and similar matching gives $(\tilde{C}_+/\tilde{C}_-) \exp(-2i\Delta
\tilde{W}) = -i$. Combining these, the quantisation condition is:
\begin{equation}
  \label{eq:34}
  m \pi = 2 \Delta \tilde{W} - \frac{\pi}{2},
\end{equation}
which is just Bohr-Sommerfeld quantisation as expected for this
simplest case.

In the range $0<\epsilon<1$, there are both regions with and without
factors $(-1)^{n/2}$, however since both boundaries have this
additional factor, it is possible to get away with Eq.~(\ref{eq:34}),
using the integral in Eq.~(\ref{eq:29}) even when
$\cos(2\tilde{W}_0^\prime)<0$.  To show that this is valid, we will
discuss the result of taking the connection at $z=z_s$ into account
explicitly.  Using the results of Sec.~\ref{sec:conn-with-allow} means that
in the connection formulae at the outer boundary, one should replace
\begin{equation}
  \label{eq:35}
  \tilde{C}_+ e^{i \Delta W} \to
  C_- 
  e^{i \left[\Delta \tilde{W}_{\text{out}} - \Delta {W}_{\text{in}}
      + {\pi N} z_s/2  \right]},
\end{equation}
and for $\tilde{C}_-$ the condition is similar, but complex conjugate.
Here, the expression $W_{\text{in}}$ in Eq.~(\ref{eq:35}) should be
understood as taking $z_i \to 0$ in the expressions following
Eq.~(\ref{eq:26}) , since there is no inner boundary for $\epsilon<1$.
With this replacement, the quantisation condition becomes:
\begin{equation}
  \label{eq:36}
  m \pi = 2 \left[\Delta \tilde{W}_{\text{out}} - \Delta {W}_{\text{in}}
      + \frac{\pi N}{2} z_s \right] - \frac{\pi}{2}.
\end{equation}
To relate the integrals $\Delta \tilde{W}_{\text{out}}, \Delta
W_{\text{in}}$ to the $\Delta \tilde{W}$ of Eq.~(\ref{eq:29}) we
should note two features.  Firstly there is a term $N (\pi/2) z_s$ in
the difference of definitions due to the inverse cosine, which is
compensated by the last term in brackets in Eq.~(\ref{eq:36}).  Other
than this, one may see that the various sign factors in front of
$\Delta W_{\text{in}}$, in the relative definition of $\Delta
W_{\text{in}}$ vs $\Delta \tilde{W}_{\text{out}}$, and in the order of
limits of integration in Eq.~(\ref{eq:26}) are such that the remaining
parts of the integrals all match  so that $\Delta{W}_{\text{in}} - \pi
N z_s/2 - \Delta \tilde{W}_{\text{out}} = \Delta \tilde{W}$.

\subsection{Above critical level}
\label{sec:above-critical-level}

Above the critical level, there can either be three or five separate
regions to consider: a forbidden region in the centre and either
single allowed regions to the left and right if $\epsilon > u$, or
pairs of allowed regions if $\epsilon < u$.  We will label these
regions, from left to right as $\tilde{L}, L, F, R, \tilde{R}$, with
the $\tilde{L},\tilde{R}$ regions having $z_s < |z| < z_o$ when
$\epsilon< u$, and vanishing otherwise.  These labels will be used
both for the coefficients $\tilde{L}_\pm, \tilde{R}_\pm$, and for the phases $\tilde{W}_{\tilde{L}}, W_L,
\Omega_F, W_R, \tilde{W}_{\tilde{R}}$.  Focusing on the right hand side ($z>0$),
it is convenient to define $W_R, \tilde{W}_R$ with the same limits as
$W, \tilde{W}$ in Eq.~(\ref{eq:26}), and to define $\Omega_F(z)$ as
the integral from $0$ to $z$.

Consider first the case $\epsilon < u$.  With the above definitions
the outer boundary condition simply becomes $\tilde{R}_+/\tilde{R}_- =
i$, as the phase in this region is defined so it vanishes at $z_o$.
Using Eq.~(\ref{eq:28}), this translates to a condition:
\begin{equation}
  \label{eq:33}
  \frac{R_+}{R_-} 
  \exp\left(i 2\left[\Delta  W_{\text{in}} - \Delta \tilde{W}_{\text{out}}
      - \frac{\pi N}{2} z_s + \frac{\pi}{2} \right]\right)
  = i,
\end{equation}
where the $\pi/2$ term on the left hand side is associated with an
overall minus sign from $i\to 1/i$ on the right hand side.

In the case $\epsilon>u$ we need only $W(z)$, which we have already
defined as the integral from $z_i$ to $z$.  In this case the phase
matching condition at the outer boundary implies:
\begin{equation}
  \label{eq:37}
  \frac{R_+}{R_-} 
  \exp\left(i 2 \Delta  W \right)
  = i, \quad
  \Delta W = \int_{z_i}^{z_o}\!\!dz [N W_0^\prime + W_1^\prime].
\end{equation}
Both Eq.~(\ref{eq:33}) and Eq.~(\ref{eq:37}) can be combined by the
statement $(R_+/R_-)\exp(2 i \Delta W_{\text{eff}}) = i$, with 
\begin{equation}
  \label{eq:43}
  \Delta {W}_{\text{eff}} =
  \begin{cases}
    \Delta  W_{\text{in}} - \Delta \tilde{W}_{\text{out}}
    - \frac{\pi N}{2} z_s + \frac{\pi}{2} & \epsilon < u \\
    \Delta W & \epsilon > u
  \end{cases}
\end{equation}
The equivalent analysis at the leftmost boundary gives
$(L_+/L_-)\exp(-2 i \Delta W_{\text{eff}}) = -i$.

The boundary at $z_i$ always involves $R_\pm$ rather than
$\tilde{R}_\pm$, and our choice of limits means that the phase is
always measured from this boundary.  Following the same logic as
led to Eq.~(\ref{eq:31}), and noting that $\gamma_i$ is
positive, one has $W(z_i + \zeta) = (2/3) |\gamma_i|^{1/2}
|\zeta|^{3/2}$ for $\zeta>0$.  At this boundary, one must match both
Airy functions, which gives the matching condition:
\begin{equation}
  \label{eq:38}
  \frac{(C_{R,b} - i C_{R,a})e^{+i\pi/4}}{(C_{R,b}+i C_{R,a}) e^{-i\pi/4}}
  =
  \frac{R_+}{R_-}
\end{equation}
where $C_{R,a}, C_{R,b}$ are the coefficients of Airy functions, as in
Eq.~(\ref{eq:18}) at the right hand ($+z_i$) boundary.  These should
also be matched to the coefficients of the growing/decaying terms in
the forbidden region.  Defining $\Delta \Omega_F = 2 \Omega_F(z_i) =
\Omega_F(z_i) - \Omega_F(-z_i)$, one may expand
\begin{math}
  \Omega_F(z_i+\zeta) = (\Delta \Omega_F/2) - (2/3) |\gamma_i|^{1/2}
  |\zeta|^{3/2}
\end{math}
for $\zeta<0$.  Matching this to Airy functions yields:
\begin{equation}
  \label{eq:40}
  \frac{C_{R,b}}{C_{R,a}} = 
  \frac{F_- e^{-\Delta \Omega_F/2}}{F_+ e^{+\Delta \Omega_F/2}}
\end{equation}
Equations~(\ref{eq:38}) and (\ref{eq:40}) can be combined to eliminate
the coefficients of Airy functions giving:
\begin{equation}
  \label{eq:39}
  \frac{%
    F_- e^{-\Delta \Omega_F/2} - i F_+ e^{+\Delta \Omega_F/2}}{%
    F_- e^{-\Delta \Omega_F/2} + i F_+ e^{+\Delta \Omega_F/2}}
  =
  \frac{R_+}{R_-} e^{-i\pi/2}
  = e^{-2 i \Delta W_{\text{eff}}}
\end{equation}
Analysis at the left hand boundary is very similar, 
\begin{equation}
  \label{eq:39}
  \frac{%
    F_+ e^{-\Delta \Omega_F/2} - i F_- e^{+\Delta \Omega_F/2}}{%
    F_+ e^{-\Delta \Omega_F/2} + i F_- e^{+\Delta \Omega_F/2}}
  =
  \frac{L_-}{L_+} e^{-i\pi/2}
  = e^{-2 i \Delta W_{\text{eff}}}
\end{equation}
Eliminating $F_{\pm}$ from these equations then yields the
quantisation condition:
\begin{equation}
  \label{eq:41}
  \frac{1+ \cos(2 \Delta W_{\text{eff}})}{1- \cos(2 \Delta W_{\text{eff}})}
  =
  e^{-2 \Delta \Omega_F}.
\end{equation}
Since the quantity $\Delta \Omega_F$ grows linearly with $N$, the
quantum tunnelling between allowed regions is hugely suppressed.
Taking this tunnelling into account at leading order yields $\cos(2
\Delta W_{\text{eff}}) \simeq -1 + 2 \exp(-2\Delta \Omega_F)$,
so that:
\begin{equation}
  \label{eq:42}
  \Delta W_{\text{eff}}(\epsilon) 
  \simeq m \pi + \frac{\pi}{2} \pm e^{-\Delta \Omega_F}.
\end{equation}
Differentiating this gives an expression for the tunnelling induced
energy splitting (the inverse of quantum tunnelling time for self
trapped states): $\delta \epsilon_{\text{splitting}} =  2e^{-\Delta
  \Omega_F} |d \Delta W_{\text{eff}}/d \epsilon|^{-1}$.

\subsection{Near critical level}
\label{sec:near-critical-level}

Near the critical level, we replace the matching to Airy functions and
the forbidden region by matching to Eq.~(\ref{eq:25}).  As in the
previous section, we either have four allowed regions or two
depending on whether $\epsilon =1$ is less than or greater than $u$.
Just as in that case, we may write the boundary conditions at the outer
boundaries as:
\begin{equation}
  \label{eq:44}
  \frac{R_+}{R_-}e^{+2i\Delta W_{\text{eff}}} = +i, \qquad
  \frac{L_+}{L_-}e^{-2i\Delta W_{\text{eff}}} = -i
\end{equation}
with $\Delta W_{\text{eff}}$ given in Eq.~(\ref{eq:42}), and $z_i \to
0$ so the phase factors vanishing at the inner boundary.  However,
because the expression for $W_1$ given in Eq.~(\ref{eq:21}) diverges
logarithmically at the critical level, it is necessary to slightly
modify the definition of $\Delta W_{\text{eff}}$, so the lower limit
of the integral for $W_1^\prime$ is taken to be $z_{\text{cutoff}}$
rather than zero, to avoid the logarithmic divergence.

Expanding Eq.~(\ref{eq:19}),~(\ref{eq:21}) near $z=0$ gives:
\begin{equation}
  \label{eq:45}
  1 - 2 (W_0^\prime)^2 = 1 - \frac{1}{2} (2u-1) z^2, \quad 
  W_1^\prime = \frac{1-\lambda}{2 |z| \sqrt{2u -1}}.
\end{equation}
Using the definitions of $\chi, \mu$ given in Eq.~(\ref{eq:24}), these expressions
can be integrated to give:
\begin{equation}
  \label{eq:46}
  W_{R(L)}(z) = \pm \frac{N \mu z^2}{4} 
  \pm \chi \ln \left(\frac{|z|}{z_\text{cutoff}}\right),
\end{equation}
for $z>0$ ($z<0$) respectively.  If we choose $z_{\text{cutoff}} =
1/\sqrt{N \mu}$, the coefficients $R_\pm, L_\pm$ can then be matched
to Eq.~(\ref{eq:25}) for $z>0$ ($z<0$).  This gives:
\begin{align}
  R_+
  &=
  \alpha \exp\left( \frac{\chi \pi}{4} + i\frac{\pi}{8} \right)
  +
  \beta \exp\left( -\frac{3 \chi \pi}{4} - i\frac{3 \pi}{8} \right)
  \\
  R_- 
  &=
  \frac{\beta \sqrt{2\pi}}{\Gamma[\frac{1}{2} - i \chi]}
  \exp\left(- \frac{\chi \pi}{4} + i \frac{\pi}{8} \right)
  \\
  L_-
  &=
  \beta \exp\left( \frac{\chi \pi}{4} + i\frac{\pi}{8} \right)
  +
  \alpha \exp\left( -\frac{3 \chi \pi}{4} - i\frac{3 \pi}{8} \right)
  \\
  L_+
  &=
  \frac{\alpha \sqrt{2\pi}}{\Gamma[\frac{1}{2} - i \chi]}
  \exp\left(- \frac{\chi \pi}{4} + i \frac{\pi}{8} \right)
\end{align}
By making use of the Weierstrass identity\cite{whittaker_watson} to
write $\Gamma[\frac{1}{2} - i\chi]\Gamma[\frac{1}{2} + i \chi] =
\pi/\cosh(\chi \pi)$,  these expressions can be combined to give the
quantisation condition:
\begin{equation}
  \label{eq:72}
  m \pi = 
  \text{Arg}\left[
    \displaystyle
    \frac{\sqrt{2\pi}}{\Gamma[1/2 + i \chi]}
    e^{2i \Delta W_\text{eff}}
    +
    e^{-\chi \pi/2} 
  \right] - \frac{\pi}{2}.
\end{equation}

\section{Conclusions and scaling of quantum break time}
\label{sec:conclusions}

We have shown how semiclassical quantisation formulae emerge from a
many-body WKB approach for the Josephson problem. We now use
this to extract the quantum break time;  this time is given by the inverse
of the anharmonicity of the energy level spacing.
At low and high energies
Bohr-Sommerfeld quantisation applies with small $1/N$ quantum
corrections. In between these two regimes, a critical level with large
quantum corrections is seen, where larger deviations from semiclassics
occur, leading to quantum break times that scale with the logarithm of
the system size.  From the quantisation formulae in the various
regimes, we can straightforwardly extract the anharmonicity of the
spectrum. For perfectly regular spacing of energy levels, the dynamics
would be periodic, matching the semiclassical dynamics. For the
critical level, the energy appears in the quantisation condition only
through $\lambda = N(\epsilon-1)$ as a prefactor appearing in $W_1$, and
through $\chi = (1-\lambda) / 2\mu$ in the gamma function in
Eq.~(\ref{eq:72}).  The former of these contributions can be written
as:
\begin{equation}
  \label{eq:47}
  \frac{d}{d\lambda} \Delta W_{\text{eff}}
  =
  - \!\!\!\!\!\!\!
  \int\limits_{z_{\text{cutoff}} = 1/\sqrt{N \mu}}^{z_o=\mu/u}\!\!\!\!
  \frac{dz}{2z \sqrt{\mu^2 - u^2 z^2}} \simeq - \frac{1}{2\mu} \ln \left[
    \frac{2 \mu \sqrt{N \mu}}{u} \right].
\end{equation}
This logarithmic dependence means that the level spacing near this
critical level is  $\delta \epsilon = \delta \lambda/N \simeq \pi / [ N d
  \Delta W_{\text{eff}}/d\lambda] \sim 1/[N\ln(N)]$, by making use of
  Eq.~(\ref{eq:47}). If this logarithm were large, then the
solutions of Eq.~(\ref{eq:72}) would require $\lambda$ to be regularly
spaced.  However, because of the $\lambda$ dependence in $\Gamma[1/2
+ i\chi]$, there will be some anharmonicity.  Since the only parameter
controlling the anharmonicity is $d\Delta W_{\text{eff}}/d\lambda$,
the anharmonicity of level spacing $\delta \delta \epsilon =
\epsilon_{n+1} + \epsilon_{n-1} - 2 \epsilon_n$ is given by $\delta
\delta \epsilon = \delta \delta \lambda/N \sim 1/[N \ln(N)^2]$.  In
contrast, away from the critical level, level spacing is $1/N$ and
anharmonicity $1/N^2$.  Thus, the quantum break time scales
logarithmically with system size near the critical level and linearly
elsewhere. This is exactly the conclusion found by considering next to
leading order corrections to the quantum dynamics via a cumulant
expansion in Ref.\cite{vardi01,anglin01} and consistent with
Ref. \cite{boukobza}.

In conclusion, we have shown how the WKB approach for many-body
systems, discussed in \cite{babelon09,keeling09:tavis-cummings} can be
applied to the dynamics of the Josephson problem, illustrating that it
can be applied to another paradigmatic problem of collective quantum
dynamics.  This many-body WKB approach is particularly useful in cases
where critical energy levels exist.  In such cases, the semiclassical
description may be inadequate even for mesoscopic systems with up to
$\sim 10^{6}$ particles, yet such numbers of particles make numerical
approaches to the full quantum dynamics very expensive.  As such, it
provides an ideal tool to identify cases where quantum dynamical
effects survive in mesocopic systems.

\begin{acknowledgments}
  JK acknowledges discussions with J. Quintanilla, and
 financial support from EPSRC grant no EP/G004714/1.
  FN acknowledges financial support from EPSRC.
\end{acknowledgments}

%\bibliography{josephson-wkb}

\end{document}